\documentclass{article}
\renewcommand{\epsilon}{\varepsilon}
\begin{document}
\title{Gas models for inflation and quintessence}

\author{
Winfried Zimdahl\footnote{Electronic address:
winfried.zimdahl@uni-konstanz.de}\\ 
Fachbereich Physik, 
Universit\"at Konstanz\\PF M678 
D-78457 Konstanz, Germany\\
Alexander B. Balakin\footnote{Electronic address: dulkyn@bancorp.ru}\\  
Department of General Relativity and Gravitation  \\
Kazan State University 420008 Kazan, Russia\\
Dominik J. Schwarz
\footnote{Electronic address: dschwarz@itp.hep.tuwien.ac.at}\\ 
Institut f\"ur Theoretische Physik, Technische Universit\"at Wien\\
Wiedner Hauptstra\ss e 3--10/136,
A-1040 Wien, Austria\\
and\\
Diego Pav\'{o}n\footnote{Electronic address:
diego@ulises.uab.es}\\ 
Departamento de F\'{\i}sica, Universidad Aut\'{o}noma de Barcelona\\
08193 Bellaterra (Barcelona), Spain
}

\date{\today}
\maketitle

\begin{abstract}
Specific internal self-interactions in gaseous cosmic fluids are shown to give rise to effective negative pressures which may violate the strong energy condition. 
On this basis we discuss the transition from  an initial de Sitter phase to a subsequent Friedmann-Lema\^{\i}tre-Robertson-Walker 
(FLRW) period as a non-equilibrium configuration of an ultrarelativistic gas. 
An accelerated expansion of the present universe is obtained as the consequence of a negative internal friction force which is self-consistently exerted on the non-relativistic cold dark matter (CDM) type microscopic constituents of the cosmic gas.
\end{abstract}

\section{Introduction}       
The currently favored models of the cosmological evolution rely on the inflationary paradigm according to which the early universe underwent a phase of accelerated expansion 
(see, e.g., \cite{LL}). 
On the other hand, there is growing evidence that the expansion of our present universe is  accelerating as well (\cite{supernovae,Riess01,boomerang}). 
Dynamically, periods of accelerated expansion are generated by an effective negative pressure of the cosmic medium. 
The usual approach to such exotic kind of matter is to resort to the dynamics of a scalar field with a 
suitable potential. 
It is the purpose of the present contribution to sketch an alternative way which uses gas models 
of the cosmic substratum.  
Gas models have the 
advantage that the relation between the microscopic dynamics 
(kinetic theory) and the macroscopic thermo-hydrodynamic description is 
exactly known (at least for simple cases). 
In particular, on this basis it is possible to calculate the macroscopic equation of state.  For standard gas models the latter is known 
to lie in the range $0< P \leq \rho/3$, where $\rho$ is the energy density and 
$P$ is the total pressure. 
Apparently, this excludes gas models as candidates for the cosmic medium 
during phases of accelerated expansion since the latter requires $P<0$. 
However, this result for the allowed equations of state relies on a dilute gas 
approximation as the basis of standard kinetic gas theory. 
This approximation is restricted to elastic, binary collisions of the particles and to close-to-equilibrium situations. 
If the dilute gas approximation is given up and additional interactions are included, the situation may change. 
Following this idea, we will regard here the cosmic medium as a self-interacting gas.  
We demonstrate, that specific self-interactions give rise to an effective negative pressure of the cosmic substratum. 
This will allow us to benefit from the mentioned advantages of gas models 
and at the same time to enlarge the range of admissible equations of state 
to negative values of 
the total pressure $P$.  In the early universe 
this procedure may generate an inflationary expansion as a specific non-equilibrium configuration 
of a self-interacting ultrarelativistic gas. 
Although such a dynamics is connected with entropy production it admits a conformal, timelike symmetry of an ``optical'' metric, which is characterized by an effective refraction index of the cosmic matter. 
In this context we discuss an exactly solvable self-consistent model of a smooth transition from a de Sitter phase to a radiation dominated FLRW period. 
We comment on the relation of this gas dynamical approach to the dynamics of scalar fields.
 
Along similar lines we investigate the present cosmic evolution in a gas dynamical framework. Here, the problem is to generate an effective negative pressure in a non-relativistic gas. 
The relevant self-interaction is shown to be equivalent to a negative internal friction force which is self-consistently exerted on CDM like constituents of 
the cosmic matter. In particular, we demonstrate that the dynamics of the 
$\Lambda$CDM model maybe recovered as a special case of cosmic ``anti-friction''. 

\section{Fluid description}

We assume the cosmic fluid to be characterized by a four velocity $u^i$ and a temperature $T$. 
In a spatially homogeneous and isotropic universe the 
Lie derivative $\pounds _{\xi }g _{ik}$ of the metric 
$g _{ik}$ with respect to the vector $\xi ^{a}\equiv  u ^{a}/T$ may be decomposed  
into contributions parallel and perpendicular to the four-velocity, 
\begin{equation}
\pounds _{\xi} g _{ik} \equiv  
\left(\frac{u _{i}}{T} \right)_{;k} 
+ \left(\frac{u _{k}}{T} \right)_{;i} = 
{\rm 2} A u _{i}u _{k}  
+ {\rm 2} \phi h _{ik} \ ,
\label{1}
\end{equation}
where 
$h ^{ik}=g ^{ik}+u ^{i}u ^{k}$.  (Units are chosen so that $c=k _{B}=1$).
The rate of change of the temperature $\dot{T}/T$, 
\begin{equation}
\frac{\dot{T}}{T }  = \frac{T}{2}u ^{i}u ^{k}
\pounds _{\xi} g _{ik} =AT \ ,
\label{2}
\end{equation}
where $\dot{T} \equiv  T _{,a}u ^{a}$, 
and the fluid expansion $\Theta \equiv  u ^{a}_{;a}$, 
\begin{equation}
\Theta \equiv 3H \equiv  3\frac{\dot{a}}{a} 
= \frac{T}{2}h ^{ik}\pounds _{_{\frac{u _{a}}{T}}} g _{ik} 
= 3T\phi \ ,
\label{3}
\end{equation}
where $H$ is the Hubble 
rate and $a$ is the scale factor of the Robertson-Walker metric,  
are given by  independent, orthogonal projections of the Lie derivative. 
Any relationship between the quantities $\dot{T}/T$ and $H$ 
represents a cooling rate of the cosmic fluid in an expanding universe.  
An important example  is   $A = - \phi $  which implies that 
$\xi^a$ is a conformal Killing vector (CKV) of the metric $g_{ik}$, 
\begin{equation}
\pounds _{\xi} g _{ik} 
= {\rm 2}\phi g _{ik}\ ,
\label{4}
\end{equation}
and, consequently, 
\begin{equation}
\frac{\dot{T}}{T} = - \frac{\dot{a}}{a}
\quad\Rightarrow\quad T \propto \frac{1}{a}\ .
\label{5}
\end{equation}
This is the well-known cooling rate of relativistic fluid (radiation) 
in an expanding universe.

Thermodynamically, the fluid is characterized by a particle flow vector 
$N ^{i}$,  
\begin{equation}
N ^{i} = n u ^{i}\ ,
\label{6}
\end{equation}
where $n$ is the particle number density, and an 
energy-momentum tensor 
\begin{equation}
T^{ik} = \rho u^{i}u^{k} 
+ P h^{ik} \ ,
\qquad P =p + \Pi \ ,
\label{7}
\end{equation}
The total pressure $P$ is the sum of an equilibrium part $p$ 
and a non-equilibrium contribution $\Pi $. 
The relevant conservation equations are    
\begin{equation}
N^{a}_{;a}  =  
\dot{n} + 3H  n =  0 \ , 
\label{8}
\end{equation}
and 
\begin{equation}
- u _{i}T^{ik}_{\ ;k} = \dot{\rho } + 3H\left(\rho + p + \Pi  \right)
= 0 \ .
\label{9}
\end{equation}
We assume equations of state in the general form 
\begin{equation}
p = p \left(n,T \right)\ ,
\mbox{\ \ }\mbox{\ \ }
\rho  = \rho  \left(n,T \right) \ ,
\label{10}
\end{equation}
i.e., particle number density and temperature are taken as the independent thermodynamical variables. 
Differentiating the latter relation for $\rho$ and using the balances (\ref{8}) and 
(\ref{9}), 
we obtain the cooling rate 
\begin{equation}
\frac{\dot{T}}{T} = - 3H \Sigma  
\left(\frac{\partial p}{\partial \rho } \right)_{n} ,
\mbox{\ \ }
\Sigma \equiv  1 - \frac{\dot{s} }{3H } = 
1 + \frac{\Pi }{nT} \ ,
\label{11}
\end{equation}
where $s$ is the entropy per particle. 
Comparing the expressions (\ref{2}) and (\ref{11}) for the temperature evolution, 
we find consistency for 
\begin{equation}
\pounds _{\xi} g _{ik}  =
\frac{{\rm 2H}}{T}\left\{g _{ik} 
+ \left[{\rm 1} - {\rm 3} {\rm \Sigma } 
\left(\frac{\partial{p}}{\partial{\rho }} \right)_{n} \right] 
u _{i}u _{k}\right\} \ .
\label{12}
\end{equation}
In the perfect fluid limit 
$\dot{s} =0$, and with $p=\rho /3$, we recover the CKV condition (\ref{4}) with
$\phi = H/T$. 
Obviously, the CKV property singles out a perfect fluid with the 
equation of state for ultrarelativistic matter. 
For any other equation of state or for $\dot{s} \neq 0$ 
the CKV condition is modified according to 
\begin{equation}
\frac{{\rm 2H}}{T}g _{ik}  \rightarrow 
\frac{{\rm 2H}}{T}\left[g _{ik} 
+ \left({\rm 1} - {\rm 3} {\rm \Sigma } \frac{\partial{p}}{\partial{\rho }} \right) 
u _{i}u _{k}\right] \ .
\label{13}
\end{equation}
The CKV condition (\ref{4}) is known to follow from Boltzmann's equation for the one-particle distribution function as an equilibrium condition for a gas of massless particles. 
The question arises to what extent the above modification of the CKV condition 
may correspond to matter with an equation of state which is different 
from $p=\rho/3$. 
To clarify this question we first introduce the concept of an 
``optical'' metric. 

\section{Optical metric}

The optical metric $\bar{g}_{ik}$ is defined by (\cite{Gordon})
\begin{equation}
\bar{g}_{ik} = g _{ik} + \left(1- \frac{1}{n _{r}^{2}} \right)u _{i}u _{k} \ ,
\label{14}
\end{equation}
where $n _{r}$ is the refraction index of the medium. Optical metrics are known to be helpful in simplifying the equations of light 
propagation in isotropic, refractive media. 
With respect to a metric $\bar{g}_{ik}$,   
light propagates as in vacuum. 
Here we are interested in the role of optical metrics 
in characterizing symmetry properties of the cosmic fluid.
Our starting point is to 
assume $\xi ^{a}\equiv  u ^{a}/T$  to be a conformal Killing vector 
of the optical metric, i.e., 
\begin{equation}
\pounds _{\xi}\bar{g}_{ik} = {\rm 2}\psi \bar{g}_{ik}
\label{15}
\end{equation}
instead of (\ref{4}).  
As a direct consequence of this assumption we find (\cite{ZBEntr}) 
\begin{equation}
\frac{\dot{n}_{r}}{n _{r}} = - \left[\frac{\dot{a}}{a} 
+ \frac{\dot{T}}{T}\right]\ .
\label{16}
\end{equation}
This means, all deviations from $aT = {\rm const}$ (corresponding to 
$\dot{s} = \Pi=0$ and $p=\frac{\rho }{3}$) 
have been mapped onto 
a time varying refraction index $n _{r}$.
Use of the temperature law (\ref{11}) provides us with  
\begin{equation}
\frac{\dot{n}_{r}}{n _{r}} = - \left[1 - 3 \frac{\partial{p}}{\partial{\rho }} \Sigma\right]\frac{\dot{a}}{a}\ .
\label{17}
\end{equation}
We are led to the statement that 
the mentioned modification of the conformal symmetry of the spacetime metric 
$g _{ik}$ (cf. Eqs. (\ref{12}), (\ref{13})) corresponds to 
the conformal symmetry of an optical metric $\bar{g}_{ik}$, characterized by an 
effective refraction index with a time dependence (\ref{17}). 
This equation establishes a relation between the rate of change of the refraction index 
and the expansion rate for a viscous cosmological fluid under the condition of a conformal symmetry of the optical metric  $\bar{g}_{ik}$.  
It is remarkable, that this conformal, timelike symmetry is compatible 
with entropy production (\cite{ZBEntr}). 

\section{Gas dynamics}    

To establish a gas dynamical background for the 
CKV condition (\ref{15}) for $\bar{g}_{ik}$  we consider  Boltzmann's equation 
for the invariant one-particle distribution function 
$ f = f\left(x,p\right)$, 
\begin{equation}
p ^{i}\frac{\partial{f}}{\partial{x ^{i}}} - \Gamma ^{k}_{il}p^{i} p^{l}
\frac{\partial{f}}{\partial{p^{k}}} 
+ \frac{\partial{\left(F ^{i}f \right)}}{\partial{p^{i}}}
= C [f,f] \ ,
\label{18}
\end{equation}
where $C$ is Boltzmann's collision integral and 
$F ^{i}$ is an effective one-particle force. 
We restrict ourselves to the class of forces which admit  solutions 
of Eq. (\ref{18}) that are 
of the type of  J\"uttner's distribution function 
\begin{equation}
f^{0}\left(x, p\right) = 
\exp{\left[\alpha + \beta_{a}p^{a}\right] } \ ,
\label{19}
\end{equation}
where $\alpha = \alpha\left(x\right)$ and 
$\beta_{a}\left(x \right)$ is timelike.  
The collision integral $C$ vanishes under such conditions, i.e., 
$C\left[f ^{0},f ^{0}\right] = 0$. 
However, there are 
contributions from $F ^{i}$ which, in general, are connected with the production 
of entropy. 
The most general expression for a force compatible with the cosmological 
principle is (\cite{ZSBP})
\begin{equation}
F ^{i} = F \left(-Ep ^{i} + m ^{2}u ^{i} \right)\ ,
\label{20}
\end{equation}
where $E \equiv - u_i p^i$ is the particle energy. 
In the special case of massless particles  
$m = 0$ this specifies to 
\begin{equation}
F ^{i} = -FEp ^{i} \qquad (m=0)\ .
\label{21}
\end{equation}
Furthermore, we assume that $F$ does not depend on $E$, i.e., 
$F = F \left(x \right)$.  
Inserting this expression for $F^i$ together with $f^0$ into (\ref{18})
and using 
the standard identifications 
\begin{equation}
\alpha =
\frac{\mu }{T}   \quad {\rm and} \quad
\beta _{a} = \frac{u _{a}}{T}\ , 
\label{22}
\end{equation}
we recover the 
conformal symmetry condition (\ref{15}) for $\bar{g}_{ik}$ 
for (\cite{ZSBP})
\begin{equation}
\left(\frac{\mu }{T} \right)^{\displaystyle \cdot} 
= \frac{3H \Pi }{nT} = 
-\dot{s} = 3 \frac{\dot{n}_{r}}{n _{r}} = 3F \ .
\label{23}
\end{equation} 
Together with (\ref{22}) the CKV condition (\ref{15}) implies the relation 
\begin{equation}
\pounds _{\xi}g _{ik} = {\rm \frac{2H}{T}}  
\left(g _{ik} 
+ \frac{\dot{s}}{3H}u _{i}u _{k} \right)\ .
\label{24}
\end{equation}
This means, we have rederived equation (\ref{12}) for $p=\rho/3$ from kinetic gas theory. 
The modification (\ref{13}) of the conformal symmetry of the spacetime metric 
$g_{ik}$ has been traced back to the action of a dissipative force (\ref{21}) 
with $F$ given by (\ref{23}). 
This force corresponds to a time dependent 
refraction index, equivalent to a time dependent chemical potential. 
In the limit $F  = \dot{s} =0$ we recover the ``global'' equilibrium properties 
$\mu/T ={\rm const}$ and the CKV relation (\ref{4}). 

The force $F^i$ depends on the microscopic particle momenta and on macroscopic fluid quantities. 
It determines the motion of the individual microscopic particles 
which themselves are the constituents of the macroscopic medium. 
Consequently, it describes a self-interaction of the fluid. 

Another aspect of this approach is worth mentioning. The symmetry requirement 
(\ref{15}) for the optical metric (\ref{14}) can only be realized if an effective 
one-particle force field $F^i$ is allowed to act on the gas particles. 
This is similar to corresponding features of gauge field theories where the requirement 
of local gauge invariance necessarily implies the existence of additional interactions, the gauge fields. 

\section{``Deflationary'' gas universe}

Now we combine the non-equilibrium fluid dynamics of the previous sections with the gravitational field dynamics. 
In a spatially homogeneous and isotropic universe Einstein's equations reduce to
\begin{equation}
3 H ^{2} = 8 \pi G \rho \ ,\quad 
\dot{H}= - 4 \pi G\left(\rho + p + \Pi  \right)\ ,
\label{25}
\end{equation}
where 
\begin{equation}
\Pi  =  - \frac{\dot{s}}{3H} p 
\label{26}
\end{equation}
according to (\ref{11}) with $p=nT$. 
For an ansatz 
\begin{equation}
\dot{s}\propto H ^{2}
\label{27}
\end{equation}
we obtain the solution 
\begin{equation}
H = 2\frac{a _{e} ^{2}}
{a ^{2} + a _{e} ^{2}}H _{e}\ .
\label{28}
\end{equation}
$H$ changes continously from $H = 2 H _{e}$, equivalent to 
$a \propto \exp{\left[Ht \right]}$ at $a=0$, to 
$H \propto a ^{-2}$, equivalent to $a \propto t ^{1/2}$, the standard radiation-dominated universe, for $a \gg a _{e}$. 
For $a<a _{e}$ we have $\ddot{a}>0$, while 
$\ddot{a}<0$ holds for $a>a _{e}$.  
The value $a _{e}$ denotes the transition from accelerated to decelerated expansion, corresponding to  
$\dot{H}_{e} = - H _{e}^{2}$. 
Originally, the solution (\ref{28}) was obtained in the framework of phenomenological vacuum decay models (\cite{LiMa}). 
Using the Hubble rate (\ref{28}) in the gas dynamical context of the 
previous section we may conclude that a force (\ref{21}) with 
\begin{equation} 
F = \frac{\dot{n}_{r}}{n _{r}} 
= - 8 H _{e}\left[\frac{a _{e}^{2}}{a ^{2} + a _{e}^{2}} \right]^{2}\ ,
\label{29}
\end{equation}
which is self-consistently exerted on the microscopic constituents  of the cosmic medium realizes the deflationary dynamics (\ref{28}). 

At the same time this transition is the manifestation of the conformal, timelike 
symmetry (\ref{15}) i.e., $u ^{a}/T$ is a CKV of the optical metric $\bar{g}_{ik}$.  
The distribution function (\ref{19}) becomes 
\begin{equation}
f ^{0} = \exp{\left[\frac{\mu }{T} \right]}
\exp{\left[-\frac{E}{T} \right]} 
= n _{r}^{3}\exp{\left[-\frac{E}{T} \right]}\ .
\label{30}
\end{equation}
Since the 
microscopic particle dynamics provides us with 
$\frac{E}{T} = {\rm const}$ (\cite{ZBEntr}), i.e., the  
equilibrium type structure is maintained, any deviations from equilibrium correspond to $\dot{n}_{r}<0$. 

So far we have established an 
exactly solvable self-consistent model 
for a cosmic gas in a specific non-equilibrium configuration which describes a smooth transition from a de Sitter to a subsequent FLRW phase. 
Moreover, this dynamics appears as the manifestation of a conformal, timelike symmetry of an associated optical metric, characterized by an effective refraction index of the cosmic   
medium (\cite{ZBEntr}).  
Similar conclusion have been obtained in a model with cosmological particle production (\cite{ZB01}). 

\section{Scalar field dynamics}

To discuss the interconnection between the gaseous fluid picture and the dynamics of a scalar field in terms of which  inflationary cosmology is usually described,  
we start with the familiar identifications 
\begin{equation}
\rho = \frac{1}{2}\dot{\phi }^{2} + V \left(\phi  \right)\ ,
\quad 
P 
= \frac{1}{2}\dot{\phi }^{2} - V \left(\phi  \right)\ ,
\label{31}
\end{equation}
or
\begin{equation}
\rho + P = \dot{\phi }^{2} \ , 
\mbox{\ \ \ \ \ \ \ \ } 
\rho - P = 2 V \left(\phi  \right)\ .
\label{32}
\end{equation}
Combining the Friedmann equation 
(\ref{25}) with the Hubble rate (\ref{28}) and (\ref{32}), we obtain 
\begin{equation}
\dot{\phi }^{2} =
4\frac{H _{e}^{2}}{2 \pi }m _{p}^{2}
\frac{a ^{2}}{a _{e}^{2}}
\left[\frac{a _{e}^{2}}{a ^{2} + a _{e}^{2}} \right]^{3} \ ,
\label{33}
\end{equation}
where $m _{p}$ is the Planck mass. 
Integration of this equation yields 
\begin{equation}
\frac{a}{a _{e}} = \sinh \left[\sqrt{2 \pi }
\frac{\phi - \phi _{0}}{m _{p}}\right]\ .
\label{34}
\end{equation}
For the potential we find (cf. \cite{MaaTayRou})
\begin{equation}
V 
= \frac{H _{e}^{2}}{2 \pi }m _{p}^{2}
\frac{3 - 2 \tanh ^{2}\left[ \sqrt{2 \pi }
\frac{\phi - \phi _{0}}{m _{p}}\right]}
{\cosh ^{4}\left[\sqrt{2 \pi }
\frac{\phi - \phi _{0}}{m _{p}}\right]}\ . 
\label{34}
\end{equation}
A scalar field description with this potential implies the same cosmological dynamics as a self-interacting gas model in which the particles 
self-consistently move under the influence of an effective one-particle force. 
The above relations allow us to change from the fluid to the scalar field picture and vice versa. 
In principle, it is possible to ``calculate'' the potential if the self-interacting fluid dynamics is known. 

\section{Cosmic anti-friction}

There is evidence from SN Ia data for our present universe to be in a 
state of accelerated expansion \cite{supernovae,Riess01}. 
This interpretation, which is indirectly also backed up by 
recent data  
from the  balloon experiments 
Boomerang and Maxima \cite{boomerang} according to which 
we live in a flat universe, requires a cosmic medium with 
sufficiently high negative pressure to violate the strong energy 
condition $\rho + 3P > 0$. Cosmic matter with a negative pressure is 
now known as ``dark energy''. 
Such type of exotic substratum is either modeled by a cosmological constant or by a scalar field called ``quintessence''. 
Based on the experience with gas models for the early universe outlined in the previous sections, it is our purpose here to find a gas dynamical alternative of the quintessential description of a present accelerated expansion of the universe. 
We assume that the observational evidence  
for an effective cosmological constant is an indication for the  existence of additional interactions  within the gaseous cosmic medium, which macroscopically 
manifest themselves as negative pressures. 
We start with a Boltzmann equation which is slightly different from (\ref{18}):  
\begin{equation}
p^{i}f,_{i} - \Gamma^{i}_{kl}p^{k}p^{l}
\frac{\partial f}{\partial
p^{i}}
 = C\left[f,f\right] + {\cal S}\left(x, p\right)  \mbox{ . } 
\label{36}
\end{equation}
Here, a scattering term ${\cal S}\left(x,p \right)$ on the right-hand side takes into account additional interactions which can not be reduced to elastic, binary interactions, described by Boltzmann's collision integral $C$.  
In particular, ${\cal S}\left(x,p \right)$ may describe production or decay processes of particles.  
For the specific structure  
\begin{equation}
{\cal S} = -m F ^{i}\frac{\partial{f}}{\partial{p ^{i}}}
\ , 
\label{37}
\end{equation}
this ``collision'' term may be taken to the left-hand side of Boltzmann's equation, resulting in (\cite{ZSBP})
\begin{equation}
p^{i}f,_{i} - \Gamma^{i}_{kl}p^{k}p^{l}
\frac{\partial f}{\partial
p^{i}} + m F ^{i}\frac{\partial{f}}{\partial{p ^{i}}}
 = C\left[f,f\right]   \mbox{ . }
\label{38}
\end{equation}
The left-hand side of this equation can be regarded as 
\[
\frac{\mbox{d}f \left(x,p \right)}{\mbox{d}\lambda }
\equiv  \frac{\partial{f}}{\partial{x ^{i}}}
\frac{\mbox{d}x ^{i}}{\mbox{d}\lambda } 
+ \frac{\partial{f}}{\partial{p ^{i}}}
\frac{\mbox{d}p ^{i}}{\mbox{d}\lambda }\ ,
\]
with 
\begin{equation}
\frac{\mbox{d} x ^{i}}{\mbox{d} \lambda  } = p ^{i}\ , \qquad
\frac{\mbox{D} p ^{i}}{\mbox{d} \lambda  } = m F ^{i}\ .
\label{39}
\end{equation}
Equations (\ref{39}) are the equations of motion for gas particles which move under 
the influence of a force field $F ^{i}=F ^{i}\left(x,p \right)$. 
The quantity $\lambda  $ is a parameter along the particle worldline which for
massive particles may be related to the proper time 
$\tau $ by
$\lambda    = \tau /m$. 
Consequently, a specific collisional interaction, described by a non-standard collision 
term ${\cal S}$, may be mapped onto an effective one-particle force $F ^{i}$. 
The freedom to move interactions from the left-hand side of Boltzmann's equation to its 
right-hand side and vice versa is a known feature in the literature (cf, e.g., \cite{Kandrup}). 
Our starting point is again the general expression (\ref{19}) for a one-particle force in a spatially homogeneous and isotropic universe. 
For the purpose of this section it is convenient to replace the function $F$ by $B = m F$. 
The non-relativistic limit in which we are interested is characterized by 
\begin{equation}
-p^i u_i \equiv E = m + \epsilon \ , \qquad 
\epsilon = \frac{m}{2}v^{2} \ll m \ .
\label{40}
\end{equation}
For the spatial projection of the force we obtain 
\begin{equation}
e_i F^i = - {B\over m}
E \sqrt{E^2 - m^2} \approx - B mv \ ,
\label{41}
\end{equation}
where $e^i u_i = 0$ and  $e^i e_i = 1$. 
This structure suggests the interpretation of $F^i$ as 
(anti-) friction force (\cite{ZSBP}). 
Macroscopically, $B\neq 0$ creates an effective pressure 
\begin{equation}
\Pi = \frac{B}{H}\rho
\label{42}
\end{equation}
in a non-relativistic gas with $p \ll \rho$. 
Obviously we have $\Pi < 0$ for $B <0$. 
The 
field equations (\ref{25}) may be written as 
\begin{equation}
\frac{H ^{\prime }}{B+H}
= \frac{3}{2(1+z)} \ ,
\label{43}
\end{equation}
where 
\[
H ^{\prime } \equiv  \frac{\mbox{d}H}{\mbox{d}z} \ ,\qquad
z = \frac{a_0}{a}  - 1 \ .
\]
To integrate  equation (\ref{43}), an assumption about $B$ is necessary. 
A simple ansatz which takes into account that for large values of 
the redshift $z$ the negative pressure has to be negligible, is $|B| \propto H^{-1}$. 
Integration of Eq.~(\ref{43}) then yields 
\begin{equation}
H = {H_0\over \sqrt{1 + \mu}}\left[\mu (1+z)^3 + 1\right]^{1/2}\ ,
\label{44}
\end{equation}
where $\mu$ is a constant and $H_0$ is the present value of the Hubble parameter. 
For $z \gg 1$ we have
\begin{equation}
H \propto (1+z)^{3/2} \ \qquad (z\gg 1)\ ,
\label{45}
\end{equation}
which is the characteristic dependence for a matter-dominated universe.
For the opposite case $z \to -1$ the Hubble rate
approaches the constant value
\begin{equation}
H \rightarrow \frac{H_0}{\sqrt{\mu + 1}} 
\qquad (z \to -1)\ .
\label{46}
\end{equation}
Consequently, the Hubble rate (\ref{44}) implies a transition 
from a
matter-dominated universe at $z\gg 1$ to a de Sitter
universe at $z \to -1$. 
In other words, 
we have recovered the $\Lambda$CDM model as a special case 
of cosmic anti-friction. 
The explicit expression for the ratio $|B|/H$ is
\begin{equation}
\frac{|B|}{H} = \frac{1}{\mu (1+z)^3 + 1} \ .
\label{47}
\end{equation}
Accelerated expansion occurs for  
$|B|/H \geq 1/3$. 
The redshift $z_{\rm acc}$ which denotes the transition from decelerated to accelerated expansion is determined by $1 + z_{\rm acc} = (2/\mu)^{1/3}$. 
For the favored value $1/(1+\mu) \approx 0.7$, accelerated expansion starts at
$z_{\rm acc}\approx 0.67$ (cf.~Ref.~\cite{ZSBP}), which is in the expected range 
$0.5 < z _{\rm acc} < 1$ (cf.~Ref.~\cite{Riess01}). 

\section{Interacting two-component model}

In this section we investigate in which sense the cosmological anti-friction scenario discussed so far may  
be regarded as an effective description for a mixture consisting of standard non-relativistic matter component and a component of non-standard, exotic matter. 
The motivation behind this study is the ``coincidence problem'', i.e., the question why the density of the dark energy is of the order of the matter density just at the present time. 
We decompose the total energy density 
\begin{equation}
\rho = {\rho_0\over \sqrt{1 + \mu}}\left[\mu (1+z)^3 + 1\right]\ ,
\label{48}
\end{equation}
where $\rho_0$ is its present value, 
into a conventional matter part (subscript M) and a
``Q matter'' part (subscript Q):
\begin{equation}
\rho = \rho _{M} + \rho _{Q}\ ,
\label{49}
\end{equation}
with 
\begin{equation}
\rho_{M} = r \rho_{0}\left(1 +z \right)^{3}
\label{50}
\end{equation}
and 
\begin{equation}
\rho_{Q} = \left[\frac{\mu (1+z)^3 + 1}
{\left(\mu + 1 \right)\left(1 +z \right)^{3}}
- r\right]
\rho_{0}\left(1 +z \right)^{3}\ ,
\label{51}
\end{equation}
where $r=r \left(a \right) > 0 $ with 
$r _{0}\equiv  r \left(a _{0} \right)<1$.
The factor $r _{0}$ fixes the ratio of both components 
at $a _{0}$: 
\begin{equation}
\frac{\rho  _{M}\left(a _{0} \right)}{\rho _{Q}\left(a _{0} \right)}
= \frac{r _{0}}{1-r _{0}}\ .
\label{52}
\end{equation} 
For $r = {\rm const}$ the universe evolves as though the cosmic 
substratum would consist of two non-interacting components: Non-relativistic 
matter with $p _{M}=0$ and Q-matter with an equation of state
\begin{equation}
P _{Q}= \Pi 
= \frac{B}{H}\rho  
\qquad \left(r= {\rm const} \right)\ .
\label{53}
\end{equation}
If $r$ is allowed to vary  we have
\begin{equation}
\dot{\rho }_{M} + 3 H \rho _{M} = \frac{\dot{r}}{r}\rho _{M}\ .
\label{54}
\end{equation}
For $\dot{r}>0$ the term on the right-hand side may be regarded as a matter
``source''. This corresponds to a ``sink'' in the energy balance of the Q 
component,
\begin{equation}
\dot{\rho }_{Q} + 3 H \rho _{Q} = -3B\left(\rho _{Q} + \rho _{M} \right) 
- \frac{\dot{r}}{r}\rho _{M}\ .
\label{55}
\end{equation}
By introducing the effective pressures
\begin{equation}
P _{M} = \Pi _{M} \equiv  - \frac{\dot{r}/r}{3H}\rho _{M}\ ,
\label{56}
\end{equation}
and
\begin{equation}
P _{Q} =  \Pi - \Pi _{M}
\ ,
\label{57}
\end{equation}
the balances (\ref{54})  and (\ref{55})  become
\begin{equation}
\dot{\rho }_{M} + 3H \left(\rho _{M} + P _{M} \right) = 0
\label{58}
\end{equation}
and
\begin{equation}
\dot{\rho } _{Q} + 3H \left(\rho _{Q} + P _{Q} \right) = 0 \ ,
\label{59}
\end{equation}
respectively. 
The quantity of interest for the coincidence problem is the ratio $\rho _{M}/ \rho _{Q}$ which is
governed by the equation
\begin{equation}
\left(\frac{\rho _{M}}{\rho _{Q}} \right)^{\displaystyle \cdot}
= 3H \left[\frac{\rho _{M}}{\rho _{Q}} \right]
\left[\frac{P _{Q}}{\rho _{Q}}
- \frac{P _{M}}{\rho _{M}}\right]\ .
\label{60}
\end{equation}
For $P _{M}=0$, or equivalently $r={\rm const}$, one has
$P _{Q}=\Pi <0$, i.e., the ratio $\rho _{M}/ \rho _{Q}$ continuously 
decreases and for large cosmological times 
$\rho _{M} \ll \rho _{Q}$ is valid. In other words, the matter component becomes
dynamically negligible.
However, if an exchange between both components is admitted, which amounts
to a nonvanishing quantity $P _{M}=\Pi _{M}$, there exists a second stationary
solution of Eq. (\ref{60}), namely
\begin{equation}
\frac{P _{Q}}{\rho _{Q}}
= \frac{P _{M}}{\rho _{M}}\ .
\label{61}
\end{equation}
This stationary solution is characterized by the value 
$r_s$ of the interaction parameter,  
\begin{equation}
r_{s} = \frac{\mu}{\left(\mu + 1 \right)^2}
\left[\mu + \left(1 +z \right)^{-3}\right]\ .
\label{62}
\end{equation}
For the corresponding asymptotic behavior of $\rho _{M}$ and $\rho _{Q}$ we 
obtain
\begin{equation}
\left(\rho _{M} \right)_{s} = 
\frac{\mu \rho _{0}}{\left(\mu + 1 \right)^{2}}
\left[1 + \mu \left(1+z \right) ^{3}\right]
\label{63}
\end{equation}
and
\begin{equation}
\left(\rho _{Q} \right)_{s} 
= \frac{\rho _{0}}{\left(\mu + 1 \right)^{2}}
\left[1 + \mu \left(1+z \right)^{3} \right]\ ,
\label{64}
\end{equation}
respectively. 
Both components red-shift at the same rate. There is a permanent energy 
transfer from the Q component to the matter component. Without transfer 
$\rho _{M}$ would red-shift as $a ^{-3}$  
while $\rho_{Q}$ would remain 
constant. The transfer makes $\rho _{Q}$ red-shift also and, on the other 
hand, it makes $\rho _{M}$  red-shift at a lower rate than without transfer.
In general, the red-shifts differ. For a specific amount of transfer, 
however, given by the expression (\ref{62}) for $r$, the rates just coincide. 
We conclude that a fixed ratio $\rho _{M}/\rho _{Q}$ is compatible with cosmic
anti-friction.
This ratio coincides with the parameter $\mu$, 
\begin{equation}
\mu  = \frac{\left(\rho _{M} \right)_{s}}
{\left(\rho _{Q} \right)_{s} }\ .
\label{65}
\end{equation}
While our considerations leave open the question, how this stationary state is approached, they suggest that an interaction between dark matter and dark energy might play an essential role for the coincidence problem (see also \cite{amendola,ZPC}). 

\section{Conclusions}

Self-interacting gas models may be useful tools to investigate the dynamics of the universe also in periods which require an effective negative pressure of the cosmic substratum. 
This has been demonstrated here both for an inflationary phase in the early universe and for a present state of accelerated cosmic expansion. 
A transition from a de Sitter phase to a FLRW period in the early universe is realized by a specific self-consistent, non-equilibrium configuration of a self-interacting relativistic gas. Although connected with entropy production, the latter admits a conformal, timelike symmetry of an optical metric with a suitable effective refraction index of the medium. 
In an analogous way we have traced back an accelerated expansion of the present universe to a self-consistent negative friction force (cosmic anti-friction) on CDM type non-relativistic gas particles. The $\Lambda$CDM model is recovered as a special case of cosmic anti-friction. 
We have demonstrated that the corresponding one-component description may be split 
into the dynamics of a two-fluid mixture
in which a Q matter component decays into CDM.
For a suitable decay rate there exists a stationary solution characterized
by a fixed, finite ratio of the energy densities of both components.  
This feature may be regarded as a hint on the possible role of an interaction between dark matter and dark energy for the cosmological dynamics.  \\
\ \\
\ \\
{\bf Acknowledgment}\\
\ \\
This paper was supported by the Deutsche Forschungsgemeinschaft,  by NATO and by the Spanish Ministry of Science and Technology under Grant BFM 2000-C-03-01. 
D.J.S. is supported by the Austrian Academy of Sciences.

\end{document}